# Numerical analysis of effective refractive index bio-sensor based on graphene-embedded slot-based dual-micro-ring resonator

C. Y. Zhao[a,b*], P. Y. Chen[a], P. Y. Li[a], C. M. Zhang[c]

[a] *College of Science, Hangzhou Dianzi University, Zhejiang, China*

[b] *State Key Laboratory of Quantum Optics and Quantum Optics Devices, Institute of Opto-Electronics, Shanxi University, Taiyuan, China*

[c] *Nokia Solutions and Networks, Hangzhou, 310053, China*

**Abstract:** We propose a novel bio-sensor structure composed of slot dual-micro-ring resonators and mono-layer graphene. Based on the electromagnetically induced transparency (EIT)-like phenomenon and the light-absorption characteristics of graphene, we present a theoretical analysis of transmission by using the coupled mode theory and Kubo formula. The results demonstrate the EIT-like spectrum with asymmetric line profile. The mode-field distributions of transmission spectrum obtained from 3D simulations based on finite-difference time-domain (FDTD) method. Our bio-sensor exhibit theoretical sensitivity of $330\,nm/RIU$, a minimum detection limit of $3.6 \times 10^{-4}\,RIU$, the maximum extinction ratio of $4.4\,dB$, the quality factor of $1.288 \times 10^3$ and a compact structure of $15\,\mu m \times 10\,\mu m$. Finally, the bio-sensor's performance is simulated for glucose solution. Our proposed design provide a promising candidate for on-chip integration with other silicon photonic element.

## 1. Introduction

Due to the large refractive index difference and transparency, silicon photonics play

Corresponding author. Email: zchy49@hdu.edu.cn.

an important role in optical integration field[1-3]. The silicon photonic devices have gained increasing research interests because of rapid development of micro-nano optical integrated fabrication technology. More importantly, the technology is compatible with complementary metal oxide semiconductor (CMOS) process. Among various silicon-based opto-electronic integrated devices, the micro-ring resonator (MRR) is one of the most noteworthy devices. Due to its small size, low loss, and no need for reflection end feedback, MRR are widely used in optical communication applications such as filter[4], modulator[5], coupler[6], and sensor[7].

In order to improve the operating performance of the integrated photonic devices, designing the innovate MRR is an effective method. Considering the slot region with large refractive index difference, the slot-based wave-guide[8] can limit the light propagating in a narrow slot region, by using slot-based wave-guide to fabricate MRR can promote effective interaction between the light field and the substance, which deduce a higher sensitivity and accuracy for the detection. A slot-based micro-ring bio-chemical sensor with a high sensitivity of $212 nm/RIU$ [9].

In order to further improve the operating performance of integrated photonic devices, taking advantage of two-dimensional material is also an effective method. Graphene has caught plenty attention due to broadband absorption, electric-optics effect and flexibly changed chemical potential. The mono-layer graphene leads to 2.3% transmittance loss for the visible light, which makes it possible to fabricate high-performance optical modulator[10]. A integrated photoelectric modulator based on mono-layer graphene covered on the ridge top can change optical absorption coefficient by controlling Fermi level[11]. A graphene-embedded ring modulator, due to its strong light-matter interaction, the shift of resonance peak is two orders of magnitude higher than that of conditional micro-ring modulators[12]. If graphene is embedded into the wave-guide[13], which shows much stronger absorption behavior due to the light-graphene interaction is greatly enhanced. Graphene-hybird MRR can obtain the electrical tunability by using a very small area of graphene. The multiple layers graphene have the stronger absorption than that of the mono-layer[14].

The EIT effect first realized in atomic systems[15]. Recently, investigation of the EIT spectrum have moved towards various coupled MRR[16-18]. We find that the optical constants can be tuned by controlling the Fermi level of graphene. The effective refractive index can be tuned by applying the gate-voltage on graphene, achieving a narrow EIT-like transmission window over a wide bandwidth[19].The EIT-like transparency window can be excellent tuned in a graphene-embedded cascaded coupled MRRs by changing the structure parameters[20].

The paper is organized as follows. In section 2, the operation principle of the MRR is illustrated by theoretical analysis and 3D FDTD simulation method. The theoretical model of graphene and electro-optics effect has an influence on resonator are explained and calculated in section 3. Finally, in section 4, the performance of MRR is showed to prove graphene make an influence on EIT-like spectrum.

**2.Theoretical model of Slot dual-MRR with graphene-embedded**

The proposed slot-based dual-MRRs is illustrated in Fig.1(a). The slot-based bus wave-guide is a conventional silicon rib wave-guide, and $2\mu m$ long mono-layer graphene is embedded into one of slot-based dual-MRR wave-guide. We put forward the most optimized design project. The system was fabricated on commerical SOI wafers, having a silicon thickness of $220\mu m$ and a buried-oxide thickness of $2\mu m$. For the critical coupling, Fig.1(b) depicts the schematic view of the radius of MRR is $5\mu m$, where the width is $270nm$, the height is $220nm$, the slot width is $75nm$, the gap between the bus wave-guide and MRR is $170nm$, the gap between two MRRs is $170nm$. The light propagation for *TE* fundamental mode. Light enters into the input port and then couples into the coupling region between the bus wave-guide and MRR. After propagating along the MRR, it couples into the other MRR and then transmit to output port, shown by the dotted line in Fig. 1(a). Only the light whose phase shift after coupling and propagating along the MRRs is integer times of $2\pi$ can occur constructive interference and resonance.

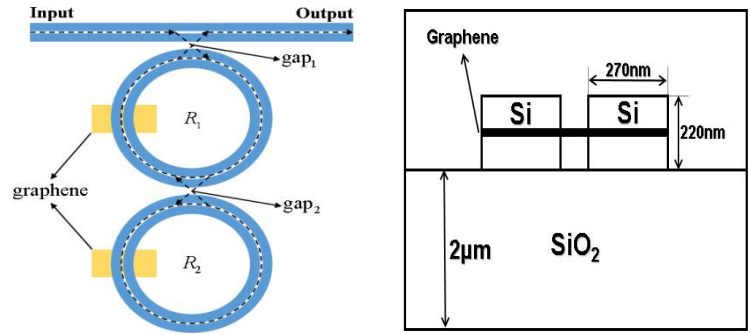

**Fig.1**. (a) The system of slot-based double MRR. (b) The section diagram of slot structure.

According to the transfer matrix method, the numerical relationship between output and input ports in coupling region can be expressed as:

$$\begin{bmatrix} E_{out1} \\ E_{out2} \end{bmatrix} = \begin{bmatrix} t & -jk \\ -jk & t \end{bmatrix} \begin{bmatrix} E_{in1} \\ E_{in2} \end{bmatrix} \quad (1)$$

Where $t = \sqrt{1-k^2}$ is self-coupling coefficient, $k$ is the cross-coupling coefficient of the

MRR.

The transmission spectrum is

$$T = 1 - \frac{(1-r_2^2)(1-a_2^2|\tau_1|^2)}{(1-r_2 a_2 |\tau_1|^2)(1+4r_2 a_2 |\tau_1|/(1-r_2 a_2 |\tau_1|^2))\sin^2[(\phi_{eff}+\phi_2)/2]} \quad (2)$$

Where $r_1$ and $r_2$ are transmission coefficients of two rings, $a_1 = e^{-\alpha_1 L_1/2}$ and $a_2 = e^{-\alpha_2 L_2/2}$ are the attenuation factor of two rings, $\alpha_1$ and $\alpha_2$ are the absorption coefficients of two rings, $R$ is the radius of ring wave-guide, $L = 2\pi R$ is the perimeter of the micro-ring wave-guide, $L_1 = L_2 = L$ is the perimeter of two ring, $\tau_1 = (r_1 - a_1 e^{i\phi_1})/(1 - r_1 a_1 e^{i\phi_1})$ is the transmission of the first ring. $\lambda$ is the incident wavelength, $n_{eff}$ is the effective refractive index of system. $\phi_1 = 2\pi/\lambda n_{eff} L_1$ and $\phi_2 = 2\pi/\lambda n_{eff} L_2$ are the phase shift induced by the two rings and $\phi_{eff} = \pi + \phi_1 + \arg((a_1 - r_1 e^{i\phi_1})/(1 - r_1 a_1 e^{i\phi_1}))$ is the equivalent phase.

The external control voltage can tune the chemical potential, and the chemical potential $\mu_c$ can control graphene absorptivity. The conductivity approach can illustrate the relationship between light absorption and chemical potential[21]. The optical conductivity of graphene $\sigma = \sigma_{intra} + \sigma'_{inter} + \sigma''_{inter}$ can be derived from Kubo formula[22]

$$\sigma_{intra} = \sigma_0 \frac{4\mu_c}{\pi (\tau_1 - i\omega)}$$

$$\sigma'_{inter} = \sigma_0 (1 + \frac{1}{\pi} tan^{-1} \frac{\omega - 2\mu_c}{\tau_2} - \frac{1}{\pi} tan^{-1} \frac{\omega + 2\mu_c}{\tau_2})$$

$$\sigma''_{inter} = \sigma_0 \frac{1}{2\pi} ln \frac{(\omega + 2\mu_c)^2 + {}^2\tau_2^2}{(\omega - 2\mu_c)^2 + {}^2\tau_2^2} \quad (3)$$

where $\omega$ is the angular frequency of the incident light, $\sigma_0 = e^2/4$ is the universal conductivity, is the reduced Plank constant. $e$ is the charge. The wavelength of incident light is $1550 nm$, external relaxation time $\tau_1 = 100 fs$, inter relaxation time $\tau_2 = 1 ps$.

The light-absorption characteristic can be illustrated by the in-plane permittivity of graphene[20]:

$$\varepsilon_g'' = \frac{e^2[1+\frac{1}{\pi}(tan^{-1}\frac{\omega-2\mu_c}{2\pi/\tau_2} - tan^{-1}\frac{\omega+2\mu_c}{2\pi/\tau_2})]}{4\omega\varepsilon_0 d} + \frac{2e^2\mu_c}{\omega\varepsilon_0 d\tau_1[(\omega)^2+(2\pi/\tau_1)^2]} \quad (4)$$

where $\varepsilon_0$ is the permittivity of vacuum and $d = 0.35 nm$ is the thickness of mono-layer graphene[23]. It can be obviously noticed that the imaginary part occurs a significant decline when $\mu_c = 0.4 eV$. The anisotropic permittivity of graphene was taken into account in the following simulation.

The transmission loss is the most important parameter to illustrate the performance of optical wave-guide. The scattering loss induced by wave-guide fabrication process, the bending loss decreases with the increasing of radius. The absorption loss $\alpha$ can be expanded as $\alpha = \alpha_s + \alpha_g$, where the silicon absorption $\alpha_s$ is invariant, and the graphene absorption $\alpha_g$ decided the value of absorption loss. At first, the imaginary part of the effective index of the graphene-silicon wave-guide is mainly decided by the Fermi level-dependent dielectric constant of graphene, which indicates the dynamic control capability of the modulation. Then, the absorption loss in the top ring is determined by the imaginary part of the effective index of the graphene-silicon wave-guide. FDTD method can accurately retrieve the transmission spectrum. By optimizing these parameters, the modulation principle of the slot dual-MRR is illustrated in Fig. 2.

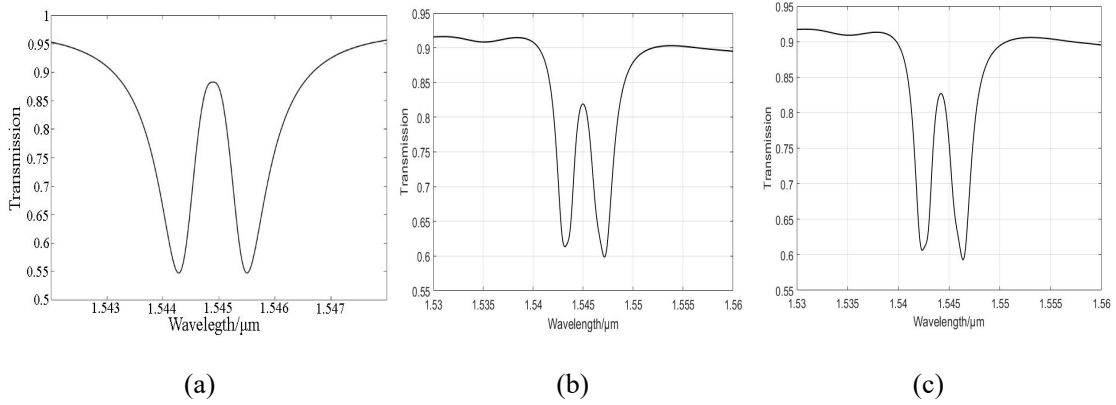

(a) (b) (c)

**Fig. 2.** (a)Numerical calculation result for the graphene embedded MRR transmission. (b)Three-dimensional FDTD simulation result for the graphene embedded MRR transmission.(c)Three-dimensional FDTD simulation result for the transmission without graphene.

We notice that the absorption light in graphene will cause a significant difference in transmission. The EIT-like phenomenon results from the destructive interference.

### 3.FDTD analsysis

The graphene's complex refractive index $n_g = \sqrt{\varepsilon} = n_g + ik_g$, where the imaginary refractive index $k_g$ determines the linear optical absorption. The light signal is strongly limited in the low refractive index slot region of slot-based dual-MRR, the nonlinear coefficient of graphene five orders magnitude higher than that of silicon[24]. In order to enlarge the nonlinear effect, insert a graphene is a more efficacious and convenient comparing with reduce the effective area $A_{eff}$. At the resonant wavelength of $\lambda_1 = 1544.97nm$, $\lambda_2 = 1543.16nm$ with the chemical potential $\mu_c = 0.4eV$.

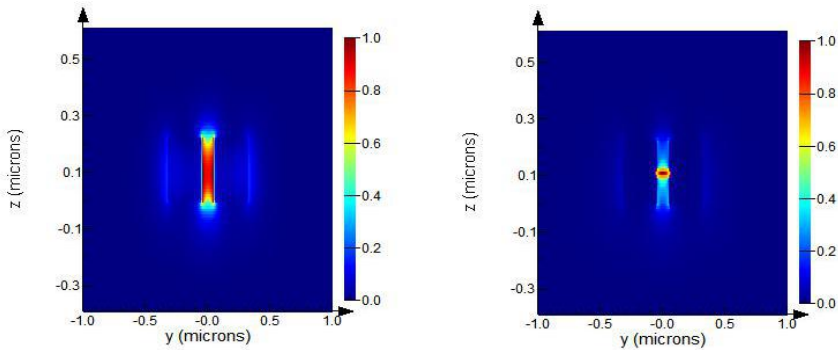

**Fig. 3.** Mode profiles (TE mode) of the MRR without graphene(a), with graphene embedded(b).

As shown in Fig.3, we can observe that the mode field is strongly confined in slot region, whatever the system is with- or without-graphene.

The mono-layer graphene embedded into the center of the micro-ring wave-guide, the interaction between graphene and optical field achieve the maximum value. It is obviously noticed that slot wave-guide structure limit the light in slot region, optical field becomes more focus on the centre part due to the absorption of graphene, which deduce more significant interaction with medium and smaller transmission loss.

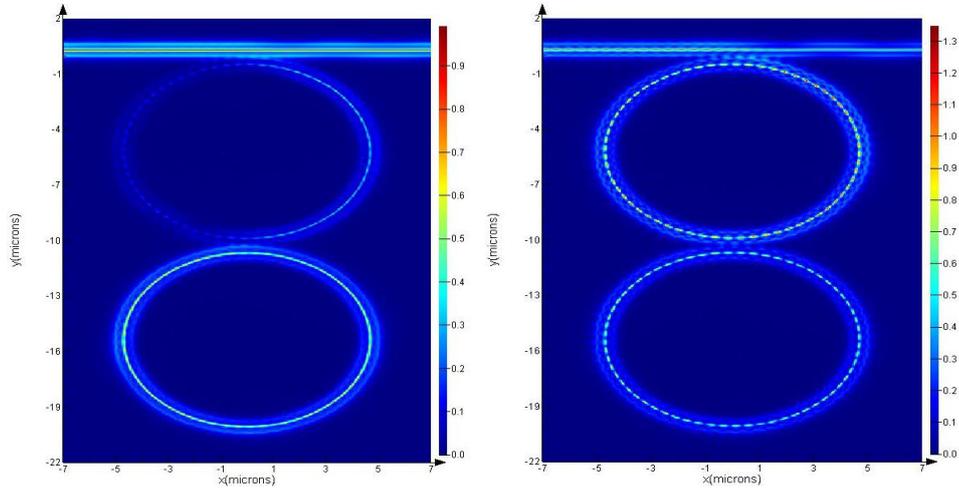

**Fig. 4.** Power profiles maps of the slot-based dual-MRR at resonant wavelengths.
(a) $\lambda_1 = 1544.97nm$, (b) $\lambda_2 = 1543.16nm$.

Based on the FDTD simulation on the power profiles of the dual-MRR. Single mode optical fiber used with wavelength $1.55\mu m$. The color bar shows the intensity of power filed has a gradually declining tendency from red to blue region. The power distribution of input port forms a strong reflection at $\lambda = 1550nm$, which is partially located at the micro-ring, namely off-resonance situation. Most of the optical power distribute is concentrated in two micro-rings at $\lambda_1 = 1544.97nm$, as shown in Fig. 4 (a). More optical power distribute localized in the second ring at $\lambda_2 = 1543.16nm$, as shown in Fig. 4 (b).

At the critical coupling condition, transmission of the dual-MRR as a function of incident light wavelength for different chemical potentials is shown in Fig.5.

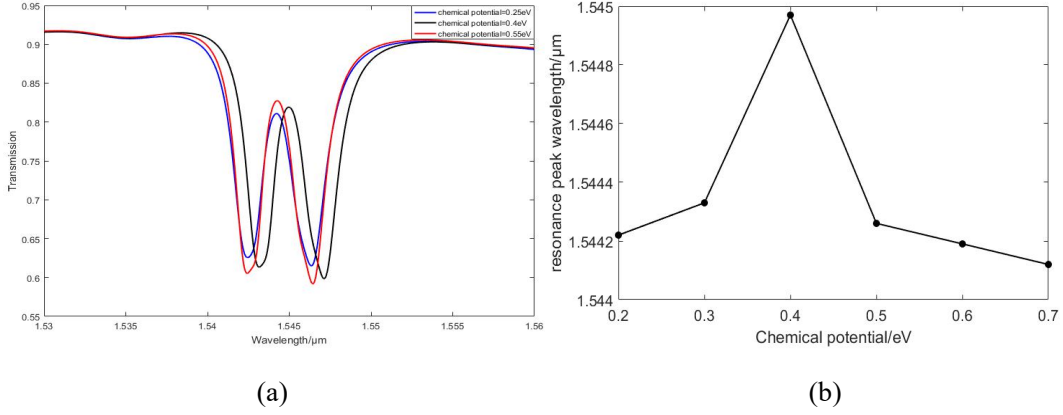

(a)                      (b)

**Fig. 5.** EIT-like spectrum of the graphene-embedded slot dual-MRR as a function of light wave-length for different Fermi level (a), the resonance wavelength as a function of different Fermi level(b).

Increasing the chemical potential from 0eV to 0.4eV, at the operation point of $\mu_c = 0.25 eV$, graphene exhibits relatively strong absorption (the blue curve). The right dip rapidly uplift and the left dip has a minimum deep. In order to show that the transparency is indeed due to a coherent cancellation, this can be shown as absorption instead of transparency. Increasing to $\mu_c = 0.4 eV$, which is the energy threshold for the photon transition in graphene. We noticed that the resonance peak wavelength exhibits a red shift first and then a blue shift. Graphene becomes transparent (the black curve), the real part showing an increase in $n_{eff}$, while the imaginary part keeping a constant, which means the light absorption of graphene strengthens gradually, resulting in the red shift. Once the chemical potential exceeds 0.4eV, the real part decreases and the imaginary part turns to zero, which means the absorption weakens and the graphene make an effect on MRR is reducing, resulting in the blue shift. Continue to increasing to $\mu_c = 0.55 eV$, the insertion loss further reduce, a small deviation from resonance lead to increasing the EIT-like transparency peak (the red curve), the right dip has maximum deep, the transmittance spectrum deforms, the left dip has the different deep as the right dip. This can be explained by the pattern of graphene's equivalent permittivity. The EIT-like effect in coupled dual-MRR systems due to photonic coherence and mode splitting. The coupling parameter $t$ in gap1 region is responsible for mode splitting, which is an crucial phenomenon occurs in coupled resonant structure[21-22].

The coherent interference between the optimize designed slot dual-MRRs will generate the EIT-like effect. The modulation of EIT-like spectrum is achieved by varying the Fermi level of graphene. As shown in Fig 6, we find that the EIT-like transparency peak will present different appearances by changing the position of mono-layer graphene from the first micro-ring to the second micro-ring.

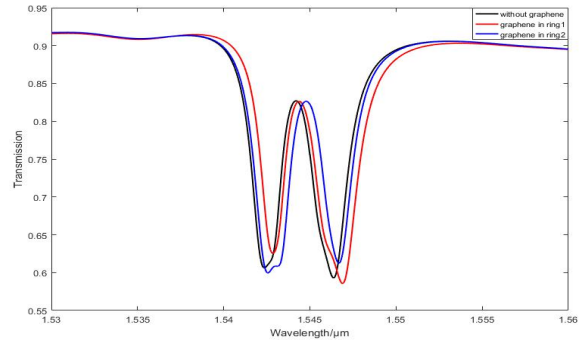

**Fig. 6.** EIT-like spectrum of the slot dual-MRR as a function of light wave-length for graphene-embedded in the first micro-ring (the red curve), graphene-embedded in the second micro-ring (the blue curve), and without graphene-embedded(the black curve)

Given the chemical potential $0.4eV$, the left peak of EIT-like spectrum deforms when the graphene-embedded in the first micro-ring, on the contrary, the right peak of EIT-like spectrum deforms when the graphene-embedded in the second micro-ring. The two lowest points of the asymmetric line-shape with a difference of magnitude.

The performance characteristics of EIT-like spectrum induced by graphene-embedded slot dual-MRRs are compared in Table1. As can be seen from the comparison of these resonators, the width of EIT-like spectrum of our resonator is much larger than those reported resonator structures, the width of EIT-like spectrum is up to $8nm$, while the radius of micro-ring only $5\mu m$. We sacrifice the depth of resonance peak, due to the big gap between bus and ring wave-guide will result in the inefficient coupling. It is necessary for us to explain whether the system under consideration in the article is experimentally designed or not. We compare some experimental results with our work in Table 1.

Table 1. Comparison of reported graphene-embedded slot dual-MRR.

| Micro-ring system | Radius | Spectrum depth | Spectrum width | Graphene position | Graphene layer | Slot structure |
|---|---|---|---|---|---|---|
| graphene-based coupled MRR[13] | $15\mu m$ | $0.8nm$ | $1nm$ | center | single | without |
| graphene-assisted ring resonator[14] | $5\mu m$ | $0.99nm$ | $1nm$ | center | single | without |
| graphene-ring resonators [20] | $20\mu m$ | $0.7nm$ | $0.3nm$ | top | single | with |
| silicon MRR wave-guide-based graphene photo-detector [23] | $1.75\mu m$ | $0.15nm$ | $1nm$ | top | single | without |
| This work | $5\mu m$ | $0.4nm$ | $8nm$ | center | single | with |

### 3. The bio-sensing analysis

The slot-based MRR can restrict light transmit into the slot region, and the absorption of graphene can effectively enhance the interaction between the light field and the cladding medium. Thus, the high sensitivity and small detection limit can be obtained. The graphene-embedded slot dual-MRR is placed in air, and the test medium fills with the slot and cladding region. The refractive index of the medium changes due to concentration, doping and other reasons, a slight increasing in the transmission peak due to the rising of the mode effective index. Taking the most extensive biochemical sensor as an example, it is very important to detect the concentration of glucose solution sensitively. The sensing principle is based on the detection of the shift in the resonant peak on account of the change in the refractive index of the upper cladding glucose solution.

The graphene-based cascaded slot MRR is placed in glucose solution of different concentrations to analyze its sensing characteristics. The change of the transmission spectrum is shown in Fig.7(a). The simulation results are shown in Fig.7(b).

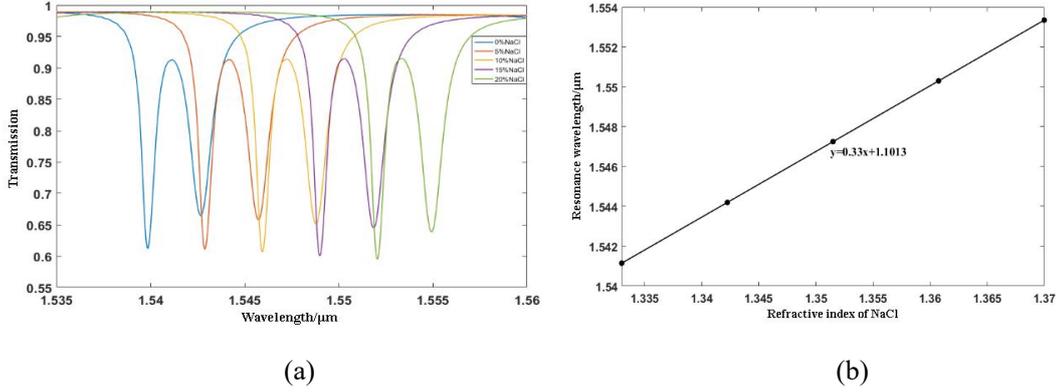

(a) (b)

**Fig. 7.** The transmission spectrum with different glucose solution.

The sensing sensitivity and minimum detection limit of the system can be expressed as [25]:

$$S = \frac{\Delta\lambda}{\Delta n_c}, \quad LOD = \frac{\Delta\lambda_{min}}{S}, \tag{5}$$

Where $\Delta\lambda$ and $\Delta n_c$ represent the change of resonance wavelength and refractive index of glucose solution, respectively, and $\Delta\lambda_{min}$ represent the change of minimum detectable resonance wavelength. The sensor sensitivity of the system is the slope of the fitting curve in Fig.7 (b) with sensitivity $S = 330\,nm/RIU$. When the concentration of glucose solution increases from 0% to 20%, the refractive index of the upper cladding region is varied from 1.333 to 1.37. The value of $\Delta\lambda = 1.8/15 = 0.12\,nm$ and $LOD = 0.12/330 = 3.64 \times 10^{-4}\,RIU$, it is the one fifteenth of $(\Delta)\text{-}3dB$ of the lowest factor that is a measurable resonance wavelength shift[26].

The performance characteristics of some integrated sensor are compared in Table 2.

Table 2. Comparison of reported some integrated sensor.

| Refractive index sensor | $S(nm/RIU)$ | $LOD(RIU)$ | $Q$-factor | Concentration |
|---|---|---|---|---|
| Slot wave-guide MRR[9] | 212 | $2.3\times10^{-4}$ | 1800 | ethanol |
| Phase shift grating MRR[22] | 297 | $1.1\times10^{-4}$ | 2000 | Sodium chloride |
| Slot wave-guide with sidewall grating[25] | 661 | $5.44\times10^{-6}$ | 37530 | — |
| Polymer MRR[27] | 188 | — | 7400 | ethanol |
| This work | 325 | $3.7\times10^{-4}$ | 1288 | glucose |

## 4. Conclusion

The slot dual-MRR enhancement the light absorption of graphene. The relationship between EIT-like spectrum and Fermi level $\mu_c$ of graphene provides the principle of modulation. The two EIT-like spectrum peaks deform independently when graphene-embedded capacitor integrated on the two different micro-rings respectively. The bio-sensor usually should be in a aqueous solution, the refractive index changes due to concentration, doping and other reasons, the changes of the effective refractive index of the MRR result in the shift of the transmission spectrum. We analyzed the sensing characteristics of graphene-embedded slot dual-MRR in different concentrations of glucose solution. The refractive index of the upper cladding region of our bio-sensor increase with the increasing of the concentration. Our bio-sensor with a compact footprint and a broad spectral response range.

**Acknowledgments**

Projected supported by the National Natural Science Foundation of China [grant number 11504074] and the State Key Laboratory of Quantum Optics and Quantum Optics Devices, Shan xi University, Shan xi, China [grant number KF201801].